\documentclass[aps,twocolumn,preprintnumbers,nofootinbib,superscriptaddress]{revtex4}
\usepackage{eurosym}
\usepackage{amsmath}
\usepackage{graphicx}
\usepackage{amsfonts}
\usepackage{array}
\usepackage{amsthm}
\usepackage{bm}
\usepackage{palatino}
\usepackage{mathpazo}
\usepackage{supertabular}
\usepackage{subfig}

\usepackage[colorlinks]{hyperref}
\usepackage{subfig}
\usepackage[font=small,labelfont=bf,justification=justified,format=plain]{caption}
\usepackage[font=small,labelfont=bf,justification=justified]{caption}
\usepackage{color}
\usepackage{booktabs}
\usepackage{multirow}
\usepackage[labelfont=bf,labelsep=quad,justification=justified]{caption}
\usepackage{xcolor}
\newcommand{\be}{\begin{equation}}
\newcommand{\ee}{\end{equation}}
\newcommand{\ba}{\begin{eqnarray}}
\newcommand{\ea}{\end{eqnarray}}
\newcommand{\bal}{\begin{align}}
\newcommand{\eal}{\end{align}}
\newcommand{\lb}{\label}

\newcommand{\bw}{\begin{widetext}}
\newcommand{\ew}{\end{widetext}}

\begin{document}

\title{Correspondence between quasinormal modes and the shadow radius in a wormhole spacetime}
\author{Kimet Jusufi}
\email{kimet.jusufi@unite.edu.mk}
\affiliation{Physics Department, State University of Tetovo, Ilinden Street nn, 1200,
Tetovo, North Macedonia}

\begin{abstract}
In this paper we study the correspondence between the real part of quasinormal modes (QNMs) and the shadow radius in a wormhole spacetime. Firstly we consider the above correspondence in a static and spherically symmetric wormhole spacetime and then explore this correspondence numerically by considering different wormhole models having specific redshift functions. To this end, we generalize this correspondence to the rotation wormhole spacetime and calculate the typical shadow radius of the rotating wormhole when viewed from the equatorial plane. We argue that due to the rotation and depending on the specific model, the typical shadow radius can increase or decrease and a reflecting point exists. Finally, we discuss whether a wormhole can mimic the black hole due to it's shadow. In the light of the EHT data, we find the upper and lower limits of the wormhole throat radius in the galactic center M87.
\end{abstract}
\maketitle

\section{Introduction}
Wormholes were first theorized by Flamm \cite{flamm}, while the famous Einstein-Rosen bridges was suggested by Einstein and Rosen \cite{Einstein35}. The spacetime topology of wormholes provides a shortcut through spacetime by connecting two different spacetime points or two universes. However Wheeler famously argued that such wormholes would be unstable and non-traversable \cite{Wheeler55}. The physics of travesable wormholes attracted considerable attention after the work of Morris, Thorne, and Yurtsever \cite{Thorne88}. An important discovery in this direction is the rotating and travesable wormhole solution first obtained by Teo  \cite{teo}. Unfortunately, the existence of wormholes is linked to the presence of exotic matter and, as a result, the energy conditions in general are violated \cite{Visser95}. 

From the astrophysical point of view, wormholes are interesting objects associated with many astrophysical phenomena, the best example is the light deflection in the strong or weak deflection limit in a given wormhole spacetime \cite{1,2,3,4,5,6,7,8,9,10,11}. It is well known that photons around the black hole can fall into a black hole or
scattered away from the black hole to infinity, yet there is the critical geodesics which separate the
first two sets, also known as unstable spherical orbits. In the observer’s sky this defines the black hole shadow \cite{Synge66,Luminet79,DeWitt73}. In this direction, it has been argued a similar effect in a wormhole spacetime, namely there is shadow boundary associated with the wormhole geometry \cite{w1,w2,w3,w4,w5,w6}. Quite amazingly, the shadow image in the galactic center of the M87 black hole was reported by the Event Horizon Telescope collaboration \cite{Akiyama1,Akiyama2}. We can therefore use these results to test the existence of different exotic objects, including the wormholes. In addition to that, general relativity predicts the existence of gravity waves. For example,  gravity waves can be produce during the collision of black holes. It is interesting that the final stage of the ringdown phase can be described in terms of the
quasinormal modes \cite{BertiCardosoWill}. The quasinormal modes (QNMs) of black holes have been investigated in many studies  \cite{Regge}-\cite{555}, but there are QNMs associated with the existence of wormholes too \cite{  Konoplya:2018ala,Churilova:2019qph,Oliveira:2018oha} 
The first evidence about the existence of gravity waves was reported by
the LIGO and VIRGO observatories \cite{AbbottBH}. We can use these observations to test not only the existence of black holes but also the possibility to distinguish black holes from wormholes or, the possibility of black hole-wormhole collision \cite{gw}. 

Interestingly, it turns out that there is a connection between the real part of the QNMs and the angular velocity of the last circular null geodesic \cite{cardoso,Hod:2017xkz,Wei:2019jve}. It was also shown that there is a correspondence between the QNMs and the strong lensing limit \cite{Stefanov:2010xz}. Recently, the author of this paper, showed a connection between the shadow radius and the real part of the QNMs in a static and a spherically symmetric black hole spacetime (see \cite{Jusufi:2019ltj}) and rotating spacetimes (see, \cite{Jusufi:2020dhz}). This correspondence was used in subsequent studies \cite{Liu:2020ola,Amir:2018pcu,hendi}, along with the analytical correspondence reported in \cite{Cuadros-Melgar:2020kqn} also used in \cite{Guo:2020nci}.  In the present work, we aim to show that such a correspondence between the real part of the QNMs and the shadow radius exists in the spacetime of static/rotating and asymptotically flat wormholes.  Such correspondence is very interesting since it relates the problem of gravity waves to the shadow. In fact, from the point of view of an observer located far away from the black hole/wormhole, we can think about the gravitational waves as massless scalar fields propagating along the last null unstable orbit and slowly leaking out to infinity. 

This paper is organized as follows. In Section II, we consider the correspondence between the shadow radius and the real part of QNMs in a static wormhole spacetime, say for electromagnetic and scalar perturbations. In Sections III, IV, and V, we consider specific models having different wormhole redshift function to study this correspondence. In Section VI, we study the observational constraints of having a wormhole in the galactic center M87 using the EFT data. In Section VII, we extend the connection between the shadow radius and the real part of the QNMs for a rotating and asymptotically flat wormholes. Finally in Section VIII, we comment on our results.  

\section{Connection between the QNMs and shadow radius in a static wormhole spacetime}
Let us start by considering a static and spehrically symmetric traversable wormhole spacetime
\begin{equation}
ds^2=-e^{2\Phi(r)}dt^2+\frac{dr^2}{1-\frac{b(r)}{r}}+r^2\left(d\theta^2+\sin^2\theta \,d\phi^2\right).
\end{equation}
We aim to study the evolution of the photon in the wormhole spacetime or the null geodesics hence we can use the  Hamilton-Jacobi equation given by
\begin{equation}
\frac{\partial S}{\partial \lambda} = - \frac{1}{2}g^{\mu\nu} \frac{\partial S}{\partial x^\mu} \frac{\partial S}{\partial x^\nu},
\end{equation}
where $\lambda$ is an affine parameter of the null geodesic.We can use the standard method for the Jacobi action $S$  separated as follows
\begin{equation}
S = \frac{1}{2} m^2 \lambda - E t + L \phi + S_{r}(r) + S_\theta (\theta),
\end{equation}
here $m$ is the mass of the particle, in our case of course for the photon we set $m=0$. Moreover $E$ and $L$ are the corresponding energy and angular momentum of the photon, respectively. Making use of the above equations we obtain \cite{w4}
\begin{eqnarray}
\frac{dt}{d\lambda} &=& \frac{E}{e^{2\Phi(r)}}, \\
\frac{e^{2\Phi(r)}}{\left(1-\frac{b(r)}{r}\right)^{1/2}}\frac{dr}{d\lambda} &=& \pm \sqrt{R(r)}, \\
r^2 \frac{d\theta}{d\lambda} &=& \pm \sqrt{\Theta(\theta)}, \\
\frac{d \phi}{d \lambda} &=& \frac{L}{r^2\sin^2\theta}.
\end{eqnarray}
with
\begin{eqnarray}
R(r) &=& E^2- \mathcal{K} \frac{e^{2\Phi(r)}}{r^2}, \\
\Theta(\theta) &=& \mathcal{K} -\frac{L}{\sin^2\theta},
\end{eqnarray}
where $\mathcal{K}$ is the Carter constant. Let us know define 
\begin{equation}
\xi = \frac{L}{E},\;\;\eta = \frac{\mathcal{K}}{E^2}.
\end{equation}

Finally if can scale $\lambda  \to \lambda E$, the radial part can be expressed in terms if the effective potential $V_{\rm eff}(r)$ as follows
\begin{equation}
 \left(\frac{dr}{d\lambda}\right)^2 + V_{\rm eff} (r)= 0,
\end{equation}
where
\begin{equation}
V_{\rm eff}(r) =-\frac{1}{e^{2\Phi(r)}} \left(1-\frac{b(r)}{r}\right)R(r),
\end{equation}
or
\begin{equation}
V_{\rm eff}(r) = -\frac{1}{e^{2\Phi(r)}} \left(1-\frac{b(r)}{r}\right)\left[1-\eta \frac{e^{2\Phi(r)}}{r^2}\right].
\end{equation}

We can determine the shadow of the wormhole by using the following conditions
\begin{equation}\lb{condition}
R(r)=0,\;\; \frac{dR(r)}{dr} =0 ,\;\;\; \frac{d^2 R(r)}{dr^2} \geq 0.
\end{equation}

Thus in our wormhole case one can show that \cite{w4}
\begin{equation}
\eta=\frac{r^2}{e^{2\Phi(r)}}|_{r_{ph}},
\end{equation}
where $r=r_{ph}$ is the radial distance of the light ring.  However as was argued in Ref. \cite{w4},
the wormhole throat acts as the position of the maximum of the potential when there are
no extrema outside the throat, the throat thus being the position of unstable circular orbits
and hence deciding the boundary of a shadow 
\begin{equation}\lb{condition}
R(r_0)=0,\;\; \frac{d^2 R(r)}{dr^2}|_{r_0} \geq 0.
\end{equation}
where $r_0$ is the wormhole throat radius. In fact, to be more precise, the throat is surely a photon sphere only if the wormhole is symmetric with respect to this throat, and the above observation of \cite{w4} is correct only in this case, while in generic asymmetric wormholes the
throat is, in general, not a photon sphere (see, \cite{Bronnikov:2018nub}). Thus, we can write
\begin{equation}
\eta=\frac{r_0^2}{e^{2\Phi(r_0)}}.
\end{equation}
In the observer’s sky this defines the wormhole shadow in terms of the celestial
coordinates 
\begin{equation}
\alpha=\lim_{r \to \infty} \left(- r^2 \sin^2\theta_0 \frac{d\phi}{dr}\right),
\end{equation}
and 
\begin{equation}
\beta=\lim_{r \to \infty} \left(r^2  \frac{d\theta}{dr}\right),
\end{equation}
with $\theta_0$ being the inclination angle. If we use the geodesic equations it follows \cite{w4}
\begin{equation}
\alpha=-\frac{\xi}{\sin\theta_0},
\end{equation}
and
\begin{equation}
\beta=\left(\eta- \frac{\xi^2}{\sin^2\theta_0}  \right)^{1/2}.
\end{equation}
For a static wormhole, the shadow radius can be found in terms of the celestial coordinates $(\alpha,\beta)$ by the simple relation
\begin{equation}\lb{RS}
R_{\rm s} = \sqrt{\alpha^2+\beta^2}=\frac{r_0}{e^{\Phi(r_0)}}.
\end{equation}

\begin{figure*}
\includegraphics[width=8.4cm]{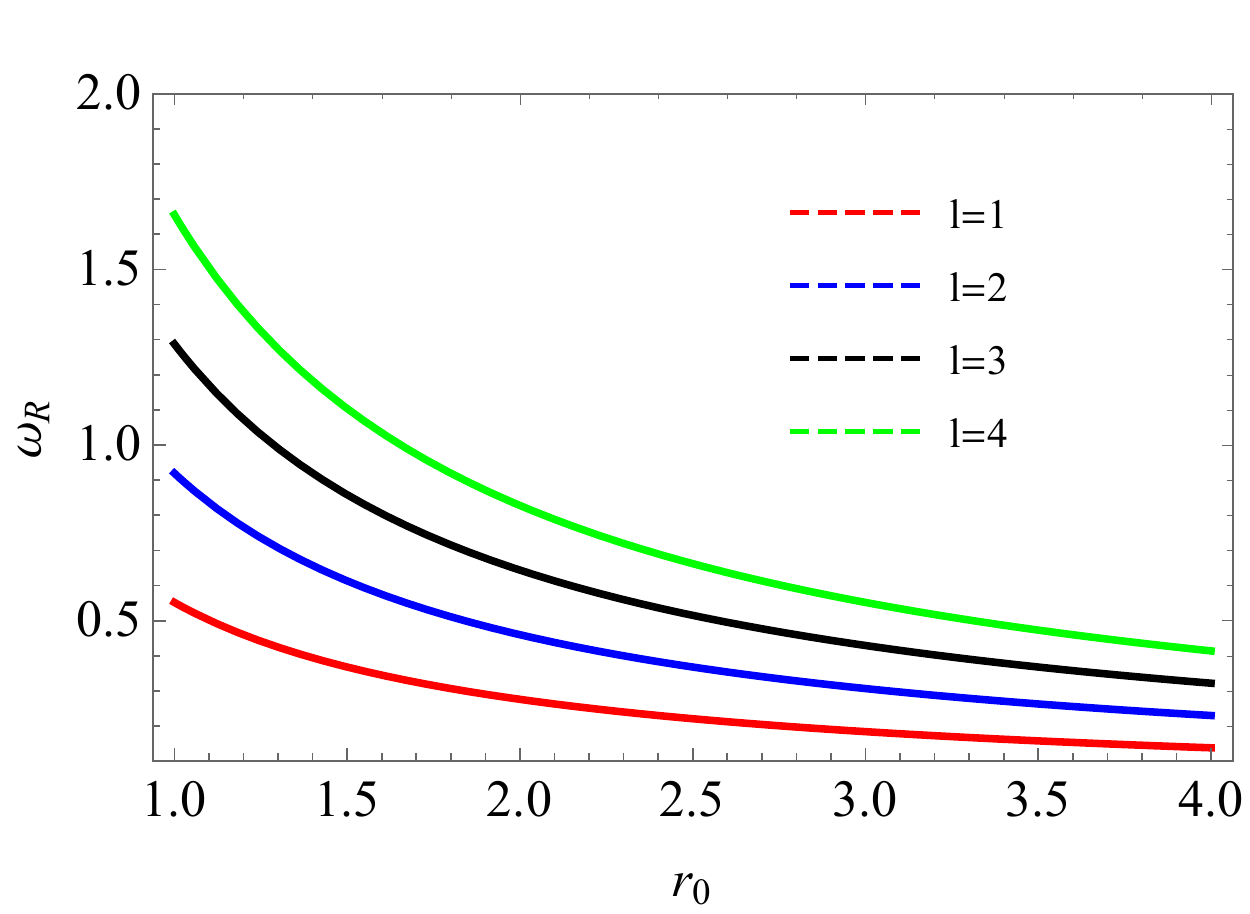}
\includegraphics[width=8.4cm]{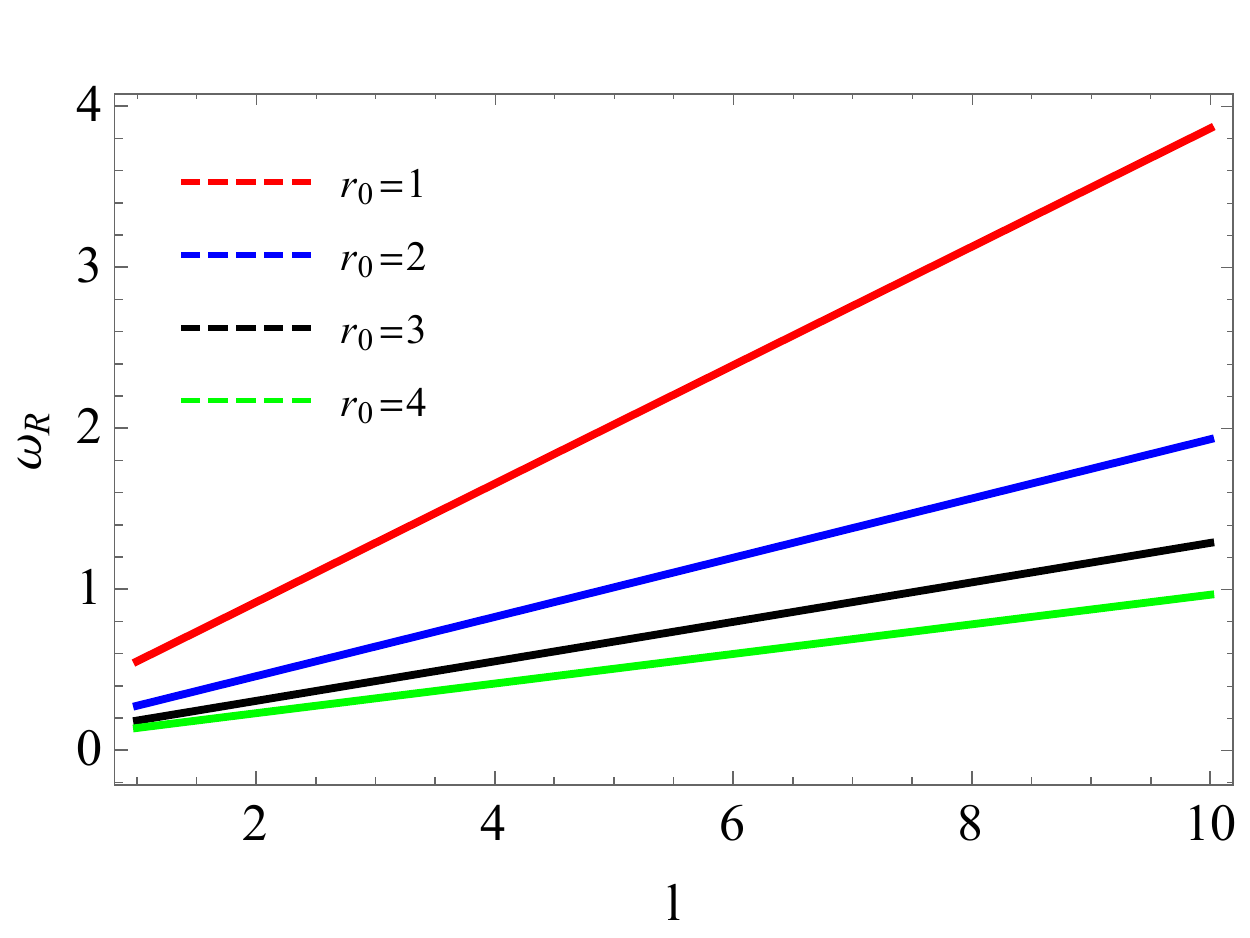}
\caption{Left panel: The real part of QNMs as a function of the wormhole throat radius $r_0$ for different values of $l$ in the case of the  model $\Phi(r)=-r_0/r$. Right panel: The real part of QNMs as a function of $l$ and fixed values of $r_0$ for the same wormhole model.  }
\end{figure*}
\begin{figure*}
\includegraphics[width=8.4cm]{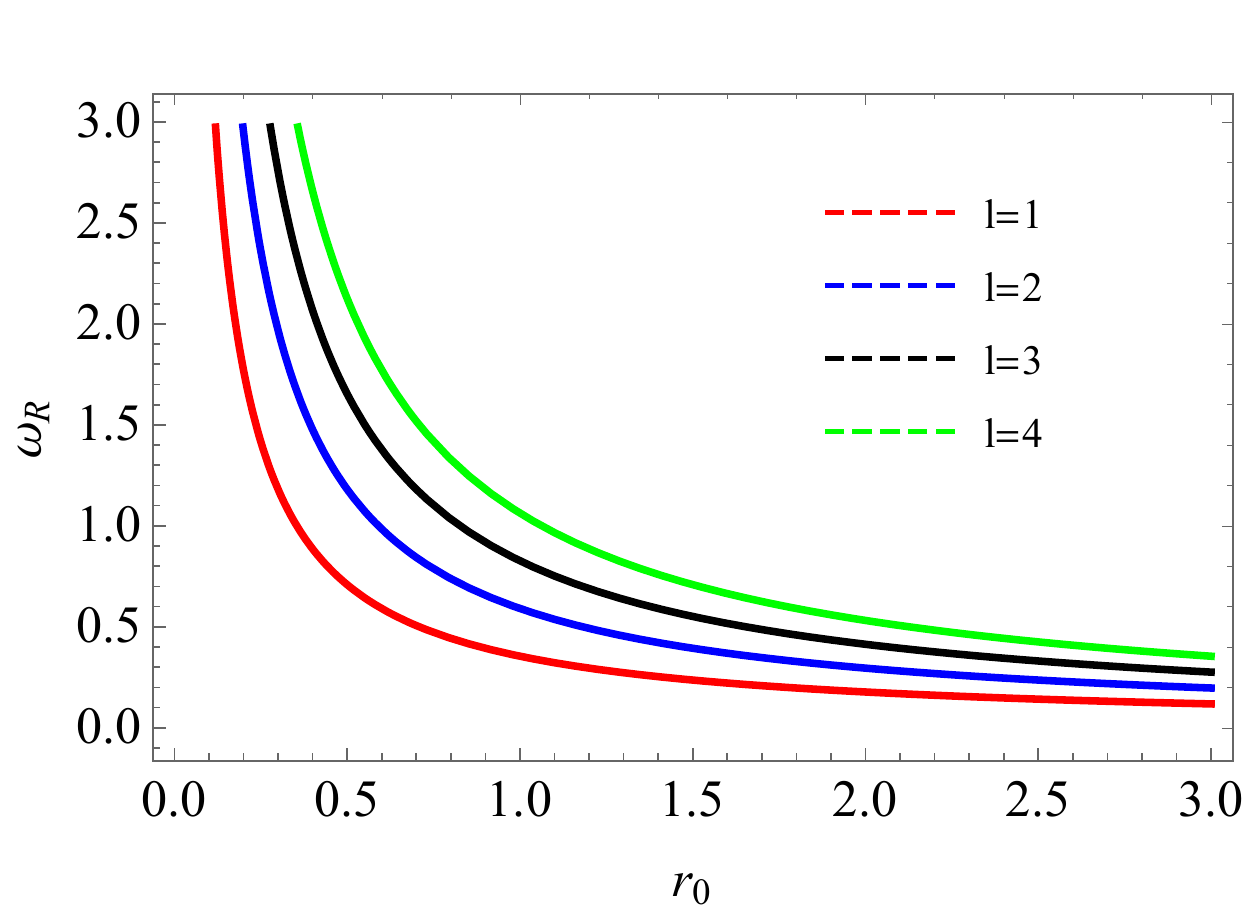}
\includegraphics[width=8.4cm]{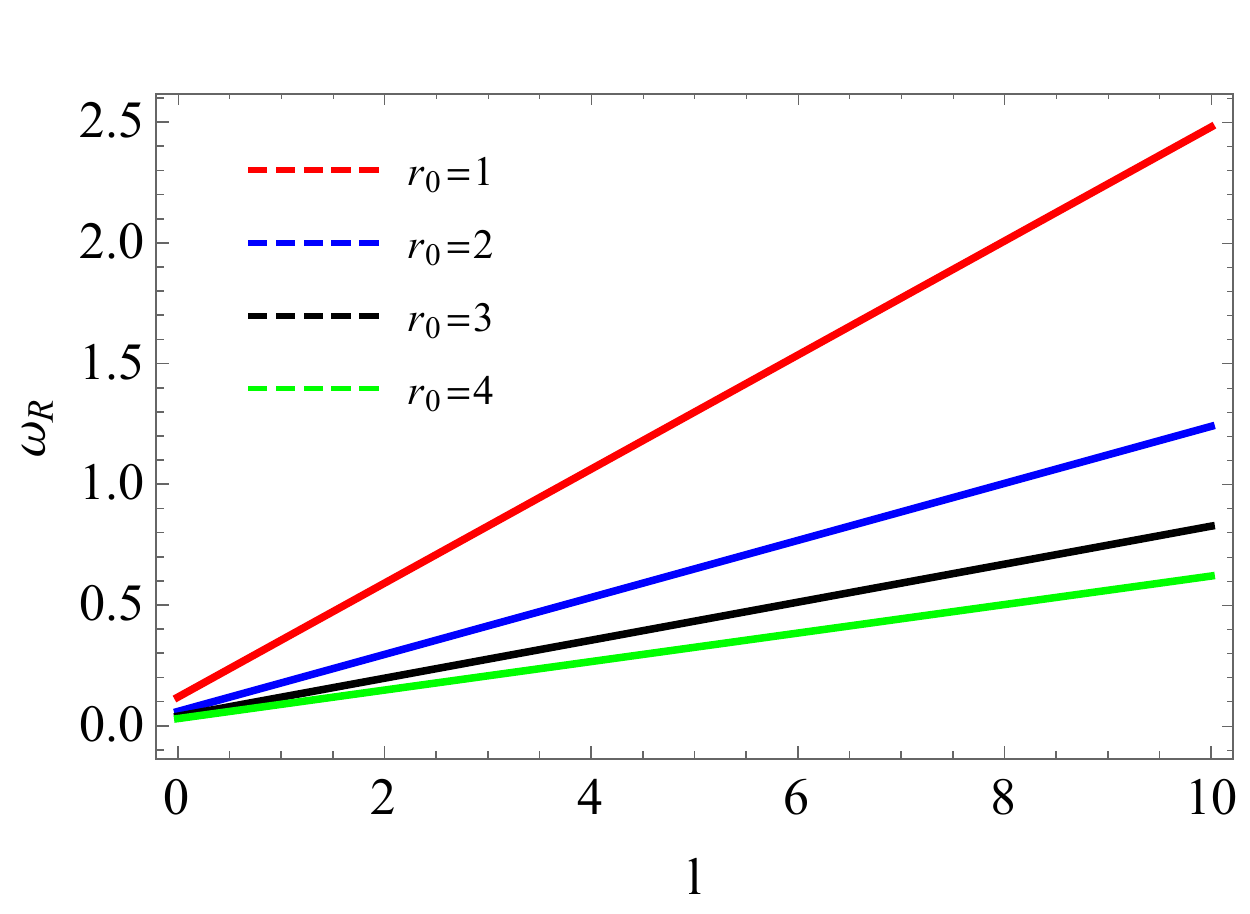}
\caption{Left panel: The real part of QNMs as a function of the wormhole throat radius $r_0$ for different values of $l$ in the case of the model $\Phi(r)=-r_0/r-r_0^2/r^2$. Right panel: The real part of QNMs as a function of $l$ and fixed values of $r_0$ for the same wormhole model. }
\end{figure*}
\begin{figure*}
\includegraphics[width=8.4cm]{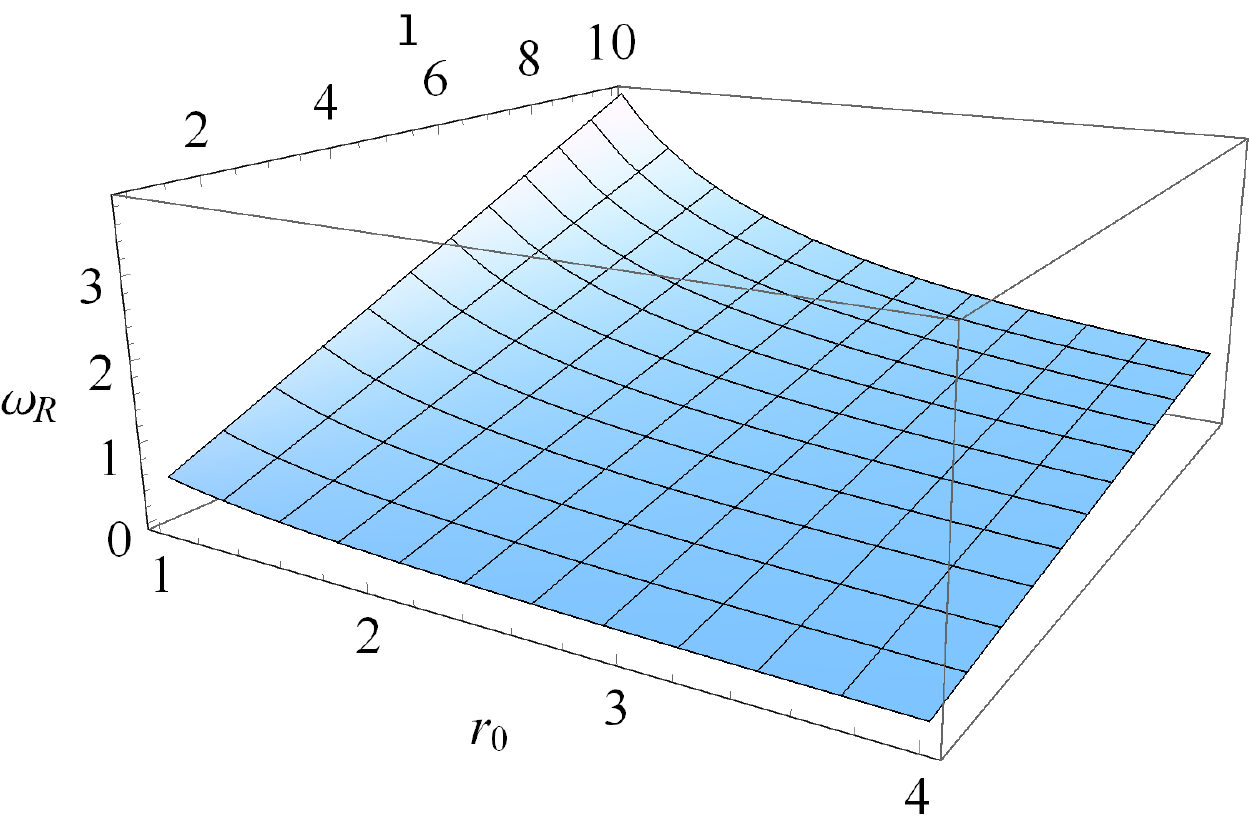}
\includegraphics[width=8.4cm]{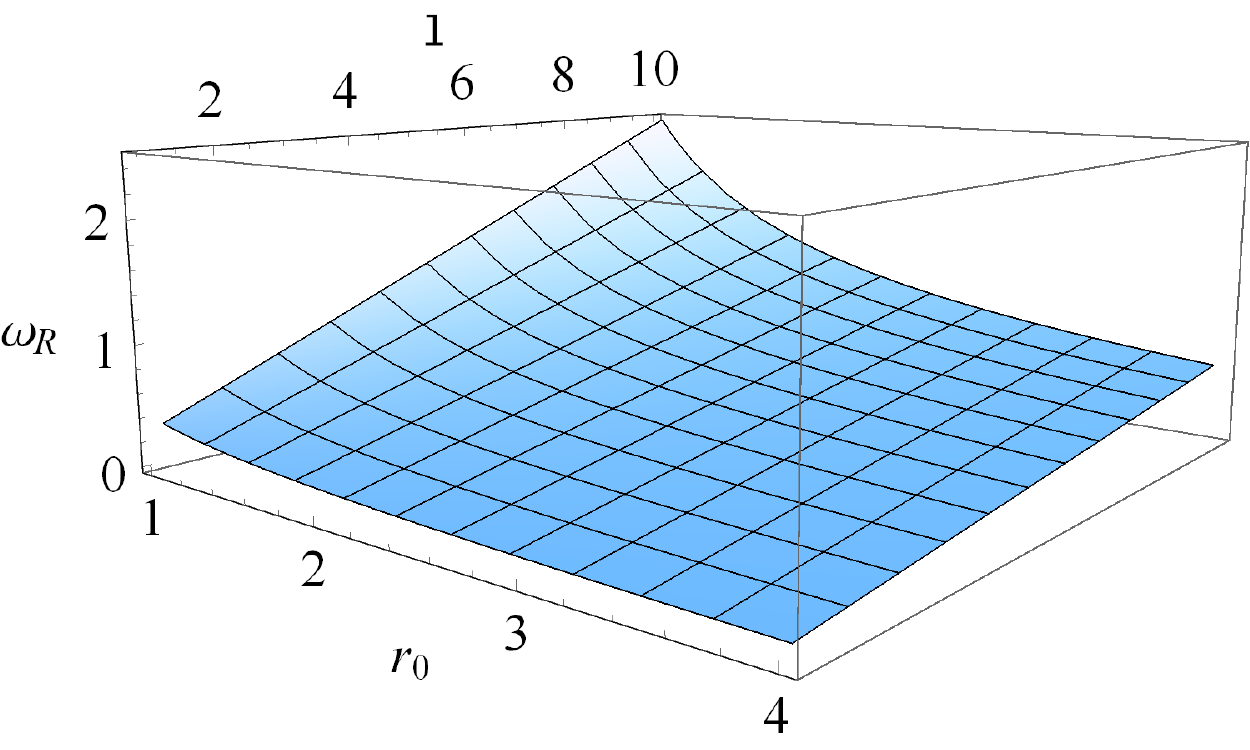}
\caption{Left panel: 3D plot of the real part of QNMs as a function of the wormhole throat radius and  $l$ using the $\Phi(r)=-r_0/r$ model. Right panel: 3D plot of the real part of QNMs as a function of the wormhole throat radius and  $l$ using the $\Phi(r)=-r_0/r-r_0^2/r^2$ model.}
\end{figure*}

\subsection{Electromagnetic field perturbations}
As we know QNMs are characteristic modes describing the final stage of a perturbed black hole or a wormhole. In the case of the black hole, one must impose an outgoing boundary condition at infinity and an ingoing boundary condition at the horizon. In the case of the wormhole, the role of horizon is played by the wormhole throat. In general, QNMs can be written in terms of the real part and the imaginary part representing the decaying modes
\begin{equation}
\omega_{QNM}=\omega_{\Re}-i \omega_{\Im}.
\end{equation}
Let us briefly discuss the electromagnetic field perturbations. To do so, we recall the wave equations for a test electromagnetic field given by
\begin{equation}
\frac{1}{\sqrt{-g}}\partial_{\nu}\left[ \sqrt{-g} g^{\alpha \mu}g^{\sigma \nu} \left(A_{\sigma,\alpha} -A_{\alpha,\sigma}\right) \right]=0,
\end{equation}
with the four-potential $A_{\mu }$. Now if we introduce the tortoise coordinate $r_{\star}$ with two branches
\begin{eqnarray}
d r_{*}=\pm \frac{dr}{e^{ \Phi(r)} \sqrt{1-\frac{b(r)}{r}}},
\end{eqnarray} 
one can obtain Schrodinger wave-like equation given by (see \cite{Konoplya:2018ala})
\begin{eqnarray}\label{schrodinger}
\frac{d^2\Psi}{dr_*^2}+[\omega^2-V(r)]\Psi=0.
\end{eqnarray}

The corresponding  effective potential is given by
\begin{equation}
V(r)=e^{2 \Phi(r)}\,\frac{l(l+1)}{r^2}.
\end{equation}

It was shown by Konoplya that one can study the shape of a wormhole by its quasinormal modes. In particular near the wormhole throat $r_0$ he obtained \cite{Konoplya:2018ala}
\begin{equation}
\omega=\frac{e^{\Phi(r_0)}}{r_0}\left(l+\frac{1}{2}\right)-i\,\left(n+\frac{1}{2}\right)\frac{e^{\Phi(r_0)}}{\sqrt{2}\,r_0}+\mathbb{O}(1/l)
\end{equation}
Just be looking at Eqs. (22) and (28) it is not difficult to see that the real part of wormhole QNMs and its shadow radius are related by 
\begin{equation}
\omega_{\Re} =\frac{1}{R_s}\left(l+\frac{1}{2}\right).
\end{equation}
However, as we already pointed out, in general, the shadow of the wormhole is determined not only by the wormhole throat, but also due to the outer light ring $r_{ph}>r_0$. So in general, the wormhole  shadow can be determined by the throat radius $r_0$, or the outer photon ring $r_{ph}$. For these reasons, we suggest that the resulting shadow radius and the real part of the QNMs are related by the same equation. 
This correspondence is normally expected to be accurate in the eikonal limit having large values of $l$ \cite{Jusufi:2019ltj}, however following the arguments in Ref. \cite{Cuadros-Melgar:2020kqn},  the values obtained by the last equation are very close even for small values of $l$. 

\subsection{Scalar field perturbations}
Alternatively, a similar result is expected if we consider a massless scalar field equation in a wormhole spacetime
\begin{equation}
\frac{1}{\sqrt{-g}}\partial_{\mu}\left(\sqrt{-g} g^{\mu \nu} \partial_{\mu}\Phi  \right)=0.
\end{equation}

The corresponding  effective potential for the scalar field is found to be (see, Ref. \cite{Churilova:2019qph})
\begin{equation}
V(r)=e^{2 \Phi(r)}\,\frac{l(l+1)}{r^2}+\frac{1}{2r}\frac{d}{dr} e^{2 \Phi(r)}   \left(1- \frac{b(r)}{r}  \right) .
\end{equation}

That being said, in the eikonal limit, the electromagnetic and the scalar field should have the same behaviour given by Eq. (28), and consequently the same correspondence between the real part of scalar QNMs and the shadow radius given by Eq. (29). We should point here that this correspondence is not guaranteed for gravitational fields, for example in the Einstein-Lovelock theory even in the eikonal limit this correspondence may be violated (see, \cite{Konoplya:2017wot}). 

\section{Model with $\Phi(r)=-\frac{r_0}{r}$}
In what follows, we are going to use different wormhole models to investigate the above correspondence. In this first model having $\Phi(r)=-\frac{r_0}{r}$, one can check that the outer photon ring and coincides with the wormhole throat, i.e. $r_0=r_{ph}$. For the shadow radius we obtain 
\begin{eqnarray}
R_s=r_0 e.
\end{eqnarray}
Hence, we find that the real part of QNMs is given in terms of the equation
\begin{eqnarray}
\omega_{\Re}=\frac{1}{r_0 e}\left(l+\frac{1}{2}\right).
\end{eqnarray}

This shows that the shadow radius of the wormhole is directly proportional to the wormhole throat radius, while the real part of the QNMs is inversely proportional to the wormhole throat radius $b_0$. We see this fact from Fig. 1 and Table I. Specifically for a fixed $l$, when we increase the wormhole throat the values of QNMs decrease and the shadow radius increases. 
\begin{table}[tbp]
\begin{tabular}{|l|l|l|l|l|l|}
\hline
\multicolumn{1}{|c|}{ } &  \multicolumn{1}{c|}{  $l=1$ } & \multicolumn{1}{c|}{  $l=2$ } &   \multicolumn{1}{c|}{  $l=3$ } & \multicolumn{1}{c|}{  $l=4$ } & \multicolumn{1}{c|}{}\\\hline
  $r_0$ &\,\,\,\,$\omega_{\Re}$  &\,\,\,\,$\omega_{\Re}$ &\,\,\,\,$\omega_{\Re}$  &\,\,\,\,$\omega_{\Re}$   & \,\,\,\,$R_s$   \\ \hline
1 & 0.5518191 & 0.9196986  & 1.2875780 & 1.6554574   & 2.718281 \\ 
2 & 0.2759095 & 0.4598493   & 0.6437890 & 0.8277287   & 5.436563 \\ 
3 & 0.1839397 & 0.3065662  & 0.4291926  & 0.5518191   & 8.154845 \\
4 & 0.1379547 & 0.2299246 &  0.3218945 & 0.4138643   & 10.87312  \\\hline
\end{tabular}
\caption{Values of the real part of QNMs and the shadow radius for the wormhole model $\Phi(r)=-\frac{r_0}{r}$. }
\end{table}
\begin{table}[tbp]
\begin{tabular}{|l|l|l|l|l|l|}
\hline
\multicolumn{1}{|c|}{ } &  \multicolumn{1}{c|}{  $l=1$ } & \multicolumn{1}{c|}{  $l=2$ } &   \multicolumn{1}{c|}{  $l=3$ } & \multicolumn{1}{c|}{  $l=4$ } & \multicolumn{1}{c|}{}\\\hline
  $r_0$ &\,\,\,\,$\omega_{\Re}$  &\,\,\,\,$\omega_{\Re}$ &\,\,\,\,$\omega_{\Re}$  &\,\,\,\,$\omega_{\Re}$   & \,\,\,\,$R_s$   \\ \hline
1 & 0.354274 & 0.590458 & 0.826641&  1.062824   & 4.234000  \\ 
2 & 0.177137& 0.295229   & 0.413320 & 0.531412  & 8.468000 \\ 
3 & 0.118091 & 0.196819 & 0.275547 & 0.354274  & 12.70200  \\\hline
\end{tabular}
\caption{Values of the real part of QNMs and the shadow radius for the wormhole model  $\Phi(r)=-\frac{r_0}{r}-\frac{r_0^2}{r^2}$.  }
\end{table}
%\end{widetext}

\section{Model with $\Phi(r)=-\frac{r_0}{r}-\frac{r_0^2}{r^2}$}
Using this model, we find that the shadow is determined by outer photon ring, i.e., $r_{ph}=2r_0$. Therefore, the shadow radius results
\begin{eqnarray}
R_s=2 r_0 e^{\frac{3}{4}},
\end{eqnarray}
which is in perfect agreement with \cite{w5}. Thus, we find for the real part of QNMs is given by
\begin{eqnarray}
\omega_{\Re}=\frac{1}{2 r_0 e^{\frac{3}{4}}}\left(l+\frac{1}{2}\right).
\end{eqnarray}

Again, we observe a rather general result, namely, the shadow radius is proportional to the wormhole throat radius, but the real part of the QNMs is inversely proportional to the wormhole throat. In Fig. 2 and in Table II, we present the plots/numerical values for the real part of QNMs by varying the wormhole throat radius and $l$. Again, for a constant $l$ an increase of $r_0$, decreases the values of the QNMs. On the other hand, having constant $r_0$, the real part of QNMs increases monotonically with the increase of $l$.

\section{Model with $\exp(2\Phi(r))=1-\frac{2 M }{r}$}
This model can be viewed as a special case of the model $\Phi(r)=-\frac{r_0}{r}-\frac{r_0^2}{r^2}$. For example one can take a series expansion around $r_0$ yielding 
\begin{equation}
\exp(2\Phi(r))=1-\frac{2M}{r}.
\end{equation}
where $M$ is a mass parameter. The wormhole shadow can be determined by the outer light ring depending on the value of the parameter $M$, since the critical orbit of the photon is given by $r_{ph}=3M$. Working in the units of $r_0=1$, we find that there is a critical value for the parameter $M=M_c=1/3$ such that, when $M \leq M_c$, the shadow is determined by the photon ring at the wormhole throat. On the other hand, when $M>M_c$, the shadow is determined from outer photon ring. In particular when $M \leq M_c$, we have the shadow radius
\begin{eqnarray}
R_s=\frac{r_0}{\sqrt{1-\frac{2M}{r_0}}}.
\end{eqnarray}
Hence, we find for the real part of QNMs in terms of the equation
\begin{eqnarray}
\omega_{\Re}=\frac{\sqrt{1-\frac{2M}{r_0}}}{r_0 }\left(l+\frac{1}{2}\right).
\end{eqnarray}
When $M > M_c$, we have the shadow radius due to the outer light ring $R_s=3 \sqrt{3} M$. Hence, we find for the real part of QNMs reads
\begin{eqnarray}
\omega_{\Re}=\frac{1}{3 \sqrt{3} M}\left(l+\frac{1}{2}\right).
\end{eqnarray}
In other words, the shadow radius in this interval becomes similar to the static black hole shadow. 
\begin{table}[tbp]
\begin{tabular}{|l|l|l|l|l|l|}
\hline
\multicolumn{1}{|c|}{ } &  \multicolumn{1}{c|}{  $l=1$ } & \multicolumn{1}{c|}{  $l=2$ } &   \multicolumn{1}{c|}{  $l=3$ } & \multicolumn{1}{c|}{  $l=4$ } & \multicolumn{1}{c|}{}\\\hline
  $M$ &\,\,\,\,$\omega_{\Re}$  &\,\,\,\,$\omega_{\Re}$ &\,\,\,\,$\omega_{\Re}$  &\,\,\,\,$\omega_{\Re}$   & \,\,\,\,$R_s$   \\ \hline
0.1 & 1.341640 & 2.236067 & 3.130495 &  4.024922   & 0.519615 \\ 
0.2 & 1.161895 & 1.936491   & 2.711088 & 3.485685   & 1.290994 \\ 
0.3 & 0.948683 & 1.581138  & 2.213594 & 2.846049   & 1.581138 \\
0.4 & 0.721687 & 1.202813 &  1.683938 & 2.165063   & 2.078460  \\
0.5 & 0.577350  & 0.962250 & 1.347150 & 1.73205  & 2.598076  \\\hline
\end{tabular}
\caption{Values of the real part of QNMs and the corresponding shadow radius for the wormhole model $\exp(2\Phi(r))=1-\frac{2M}{r}$. We have set $r_0=1$. }
\end{table}
\begin{table}[tbp]
\begin{tabular}{|l|l|l|l|l|l|}
\hline
\multicolumn{1}{|c|}{ } &  \multicolumn{1}{c|}{  $l=1$ } & \multicolumn{1}{c|}{  $l=2$ } &   \multicolumn{1}{c|}{  $l=3$ } & \multicolumn{1}{c|}{  $l=4$ } & \multicolumn{1}{c|}{}\\\hline
  $r_0$ &\,$\omega_{\Re}$  &\,$\omega_{\Re}$ &\,$\omega_{\Re}$  &\,$\omega_{\Re}$   & \,$R_s$   \\ \hline
1 & 1.50 & 2.50 & 3.50 &  4.50   & 1  \\ 
2 & 0.75 &  1.25   & 1.75  & 2.25  & 2 \\ 
3 & 0.50 & 0.83 & 1.16 & 1.50  & 3  \\\hline
\end{tabular}
\caption{Values of the real part QNMs and the shadow radius for the wormhole model  $\Phi(r)=0$.  }
\end{table}

We present our numerical estimations for the QNMs and shadow radius in Table III. It interesting to note that the value of the mode $l=1$ is given by $\omega_{\Re}=1.34$, which is very close to the value obtained via the WKB method $\omega_{\Re}= 1.35$ (see for example Ref. \cite{Churilova:2019qph}).

\section{Model with $\Phi=0.$}
This is the simplest wormhole model and the shadow radius reads
\begin{eqnarray}
R_s=r_0.
\end{eqnarray}
The real part of QNMs is therefore given by
\begin{eqnarray}
\omega_{\Re}=\frac{1}{r_0 }\left(l+\frac{1}{2}\right).
\end{eqnarray}

In Table IV we present our numerical values for the QNMs. The values of the real part of the QNMs and the shadow radius are quite simply obtained, for example consider the case $r_0=1$, we obtain: $\omega_{\Re}=1.5$ for the mode $l=1$, $\omega_{\Re}=2.5$ for the  mode $l=2$ and $\omega_{\Re}=3.5$ for the mode $l=3$. These values are very close to those obtained via the WKB approximation (see for example Ref. \cite{Oliveira:2018oha}) [$\omega_{\Re}=1.48$ for the mode $l=1$, $\omega_{\Re}=2.49$ for the mode $l=2$, and $\omega_{\Re}= 3.49$ for the mode $l=3$, respectively].

\section{Observational constraints}
\subsection{Model with $\Phi(r)=-\frac{r_0}{r}$} 
Here we shall consider the possibility of having a traversable wormhole in the galactic center M87.  We can use the reported angular size of the shadow in the M87 center detected by the EHT  $\theta_s = (42 \pm 3)\mu as$, along with the distance to M87 given by  $D = 16.8 $ Mpc, and the
mass of M87 central object $M = 6.5 \times 10^9$ M\textsubscript{\(\odot\)}. In order to constrain the wormhole throat radius $r_0$, for simplicity,  we are going to neglect the rotation.  The diameter of the shadow in units of mass $d_{M87}$ given by \cite{Allahyari:2019jqz,Khodadi:2020jij}
\begin{eqnarray}
d_{M87}=\frac{D \,\theta_s}{M}=11.0 \pm 1.5.
\end{eqnarray}

Within $1\sigma $ confidence we have the interval $9.5 \leq d_{M87} \leq 12.5$, whereas within $2\sigma $ uncertainties we have $8 \leq d_{M8} \leq 14 $ \cite{Allahyari:2019jqz,Khodadi:2020jij}. In Fig. 4 (left panel) we show the regions of parameter space of the diameter of the shadow and the throat radius $r_0$ using the wormhole model $\Phi(r)=-\frac{r_0}{r}$. Within $1\sigma$ confidence, we find the upper and lower limits of the wormhole throat radius $1.74 \leq  r_0 \leq 2.29$. On the other hand, within $2\sigma$ confidence, we find the interval $ 1.47 \leq r_0 \leq 2.57$.  Notice that $r_0$ is measured in the units of the M87 object mass $M$, therefore the expected value of the wormhole throat can be also given by $r_0 \sim (4.6-8.1) \times 10^{-4}$ pc within $2\sigma$.  This shows that the detected diameter of the shadow in the central region of the M87 galaxy could be a wormholes an a wormhole can mimic the supermassive black hole. Such values are to be expected.

\subsection{Model with $\Phi(r)=-\frac{r_0}{r}-\frac{r_0^2}{r^2}$} 
In this particular model, within $1\sigma $ and $2\sigma$ uncertainties, we the upper and lower limits of the wormhole throat radius $1.12 \leq  r_0 \leq 1.47 $ and  $0.94 \leq r_0 \leq 1.65$, respectively. Alternatively, the wormhole throat is found to be $r_0 \sim (2.9-5.2) \times 10^{-4}$ pc within $2\sigma$.

\begin{figure*}
\includegraphics[width=8.4cm]{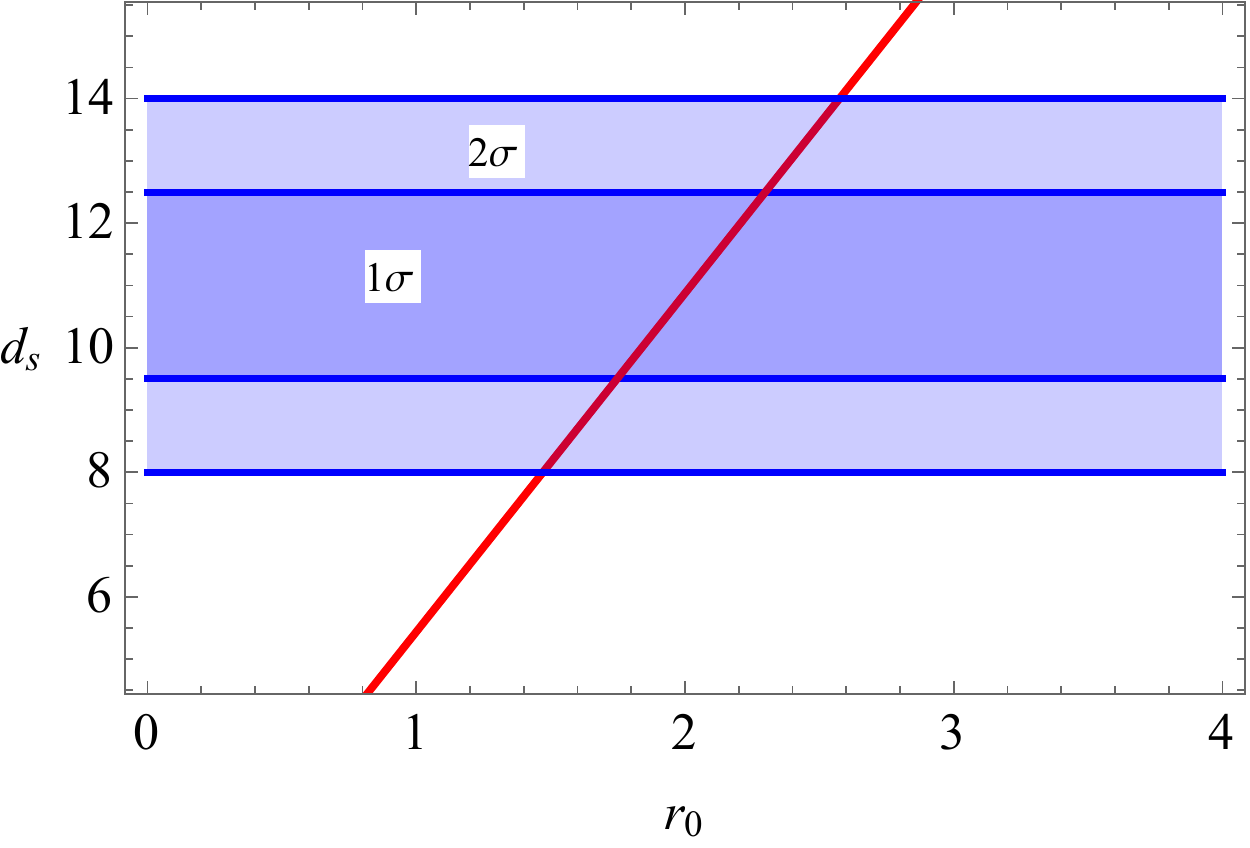}
\includegraphics[width=8.4cm]{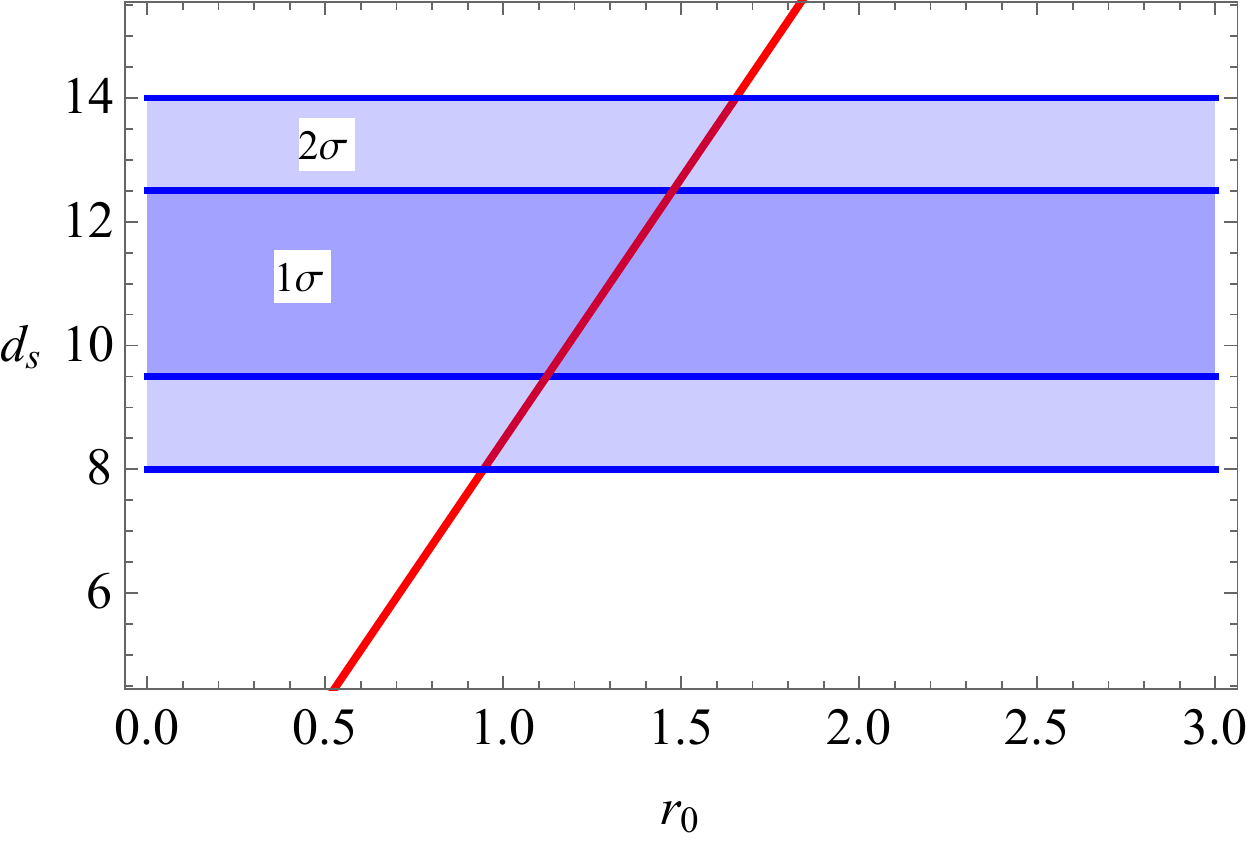}
\caption{Left panel: The regions of parameter space of the diameter of the shadow and the throat radius $r_0$ using the wormhole model $\Phi(r)=-r_0/r$ within $1\sigma $ and $2\sigma$ uncertainties, respectively. Right panel: The regions of parameter space of the diameter of the shadow and the throat radius $r_0$ using the wormhole model $\Phi(r)=-r_0/r-r_0^2/r^2$ within $1\sigma $ and $2\sigma$ uncertainties, respectively. }
\end{figure*}
%\begin{widetext}

%\subsection{Model with $\exp(2\Phi(r))=1-\frac{2M}{r}$} 

\section{Connection between shadow radius and QNMs in a rotating wormhole spacetime \label{seccon}}

Let us examine the correspondence in the rotating wormhole spcetime. To do so, we nned to write the Teo wormhole metric which describes a rotating wormhole spacetime given as follows \cite{teo,w4,w5}
\begin{equation} \label{1}
\mathrm{d}s^2=-N^2 \mathrm{d}t^2+\frac{\mathrm{d}r^2}{\left(1-\frac{b_0}{r}\right)}+r^2 K^2\left[\mathrm{d}\theta^2+\sin^2 \theta \left(\mathrm{d}\varphi-\omega \mathrm{d}t\right)^2\right]
\end{equation}
where $N= e^{\Phi(r)}$ is the redshift function which should be finite and nonzero. A particular choice of the metric functions frequently used in the literature is
given by
\begin{eqnarray}
%N&=&1-\frac{M}{r}+\mathcal{O}(1/r^2)\\
%K&=&1+\mathcal{O}(1/r)\\
\omega &=& \frac{2 J}{r^3}+\mathcal{O}(1/r^4),
\end{eqnarray}
where $J$ its angular momentum. In the following discussion, we shall use the spin parameter of the wormhole, defined as $a = J/M^2$, as a measure of the rotation rate, where $M$ is the mass of the wormhole. Without loss of generality, we are going to set the wormhole mass to unity i.e, $M=1$ [hence, we have $a=J$] along with $K=1$.
%\begin{eqnarray}
%N&=& e^{\Phi(r)}= \exp\left[-\frac{r_0}{r}-\alpha \left(\frac{r_0}{r}\right)^{\delta} \right],\,\,\ K=1,\\
%b(r)&=& r_0 \left(\frac{r_0}{r}\right)^{\gamma}.
%\end{eqnarray}
Furthermore, $r$ is a positive with the range of the radial coordinate $r \geq r_0$. The throat of the wormhole is at $b_0$ [or in our notation $r_0$] with the flare-out condition 
\begin{equation}
\frac{r_0-r_{0,r}r}{2 r_0^2}>0.
\end{equation}

In the case of vanishing spin angular momentum i.e. $a=0$, the Teo wormhole metric reduces to the static wormhole spacetime. The corresponding rotating wormhole metric in the coordinates $(t,r,\theta,\phi)$ [in the equatorial plane having set  $\theta=\pi/2$], is given by
\begin{equation}\label{met}
	{\rm d}s^2|_{\theta=\pi/2}=g_{tt}{\rm d}t^2+2 g_{t \phi}{\rm d}t {\rm d}\phi+g_{rr} {\rm d}r^2+
	g_{\phi \phi}{\rm d}\phi^2,
\end{equation}
where 
\begin{eqnarray}
g_{tt} &=& -\left(e^{2\Phi(r)}-\frac{4a^2}{r^4}\right),\\
g_{t \phi} &=& -\frac{2 a}{r},\\
g_{rr} &=& \frac{1}{1-\frac{b_0}{r}}, \\
g_{\phi \phi} &=& r^2.
\end{eqnarray}

The Lagrangian can be written as
\begin{equation}
\mathcal{L}=\frac{1}{2}\left(g_{tt}\dot{t}^2+g_{rr} \dot{r}^2+2 g_{t \phi} \dot{t} \dot{\phi}+g_{\phi \phi}\dot{\phi}^2\right).
\end{equation}

Due to the symmetry of the spacetime we have two constants of motions obtained via  the generalized momenta 
\begin{eqnarray}
p_t&=&g_{tt}\dot{t}+g_{t \phi} \dot{\phi}=-E\\
p_{\phi}&=&g_{t\phi}\dot{t}+g_{\phi \phi} \dot{\phi}=L\\
p_r&=&g_{rr}\dot{r}
\end{eqnarray}

If we now define the Hamiltonian as follows
\begin{equation}
\mathcal{H}=p_t \dot{t}+p_{\phi }\dot{\phi}+p_{r} \dot{r}-\mathcal{L},
\end{equation} 
along with the additional condition for the existence of circular geodesics at $r=r_*$, and stated as
\begin{equation}
\label{Vreq}
\mathcal{V}_r=\mathcal{V}^{'}_{r}=0.
\end{equation}
Following \cite{Jusufi:2020dhz}, we define the following quantity
\begin{equation}
R_s=\frac{L}{E}.
\end{equation}
As a result we obtain from the first condition
\begin{equation}
r^2|_{r=r_*}-e^{2\Phi(r)}R_s^2|_{r=r_*}-\frac{4 a  R_s}{r}|_{r=r_*}+\frac{4 a^2 R_s^2}{r^4}|_{r=r_*}=0,
\end{equation}
along with a second equation  
\begin{equation}
r|_{r=r_*}-\Phi'(r)e^{2\Phi(r)}R_s^2|_{r=r_*}+\frac{2 a R_s}{r^2}|_{r=r_*}-\frac{8 a^2 R_s^2}{r^5}|_{r=r_*}=0,
\end{equation}
from the second condition, respectively. As was argued in Ref. \cite{w4,w5}, the shadow boundary of a rotating wormhole is determined by the superposition of the two curves, namely a curve due to the inner light ring located at the
wormhole throat $r_0$, and the outer one located at some radial distance $r_* >r_0$.  Let us consider two  special cases for the rotating wormhole having different redshift functions.

 \subsection{Model with $\Phi(r)=-\frac{r_0}{r}$}

To determine these points let us first solve Eq. (58) for $R_s$, yielding
\begin{equation}\label{Eq34}
R_s^{\pm}=\frac{r^3}{\pm r^2\,e^{-\frac{r_0}{r}}+2a}|_{r=r_*^{\pm}}.
\end{equation}

The quantity $R_s^{\pm}$ has units of mass and the spin parameter $a$ has the units of $M^2$.  The shadow radius of a rotating black hole or wormhole depends on the observer's viewing angle $\theta_0$. In the the case with $\theta_0=\pi/2$, we can adopt the definition for the typical shadow radius which can be written as \cite{Feng:2019zzn,Jusufi:2020dhz}
\begin{equation}
\bar{R}_s=\frac{1}{2}\left(R^+_s|_{r_*^+}-R^-_s|_{r_*^-}\right)
\end{equation}
where the points $r_0^{\pm}$ are determined from Eq. (59). In this model we find  
\begin{equation}
r|_{r=r_*^{\pm}}-\frac{{R_s^{\pm}}^2 r_0}{r^2 e^{ \frac{2r_0}{r}}}|_{r=r_+^{\pm}}+\frac{2 a R_s^{\pm}}{r^2}|_{r=r_*^{\pm}}-\frac{8 a^2 {R_s^{\pm}}^2}{r^5}|_{r=r_*^{\pm}}=0,
\end{equation}

 From the last equation one can check that there is a contribution at the the outer one with $r_*^- > r_0$ only from the branch $R_s^-$. On the other hand, for the branch $R_s^+$, there is a contribution due to the wormhole throat i.e., $r_*^+=r_0$. The typical shadow radius yields
\begin{equation}
\bar{R}_s=\frac{r^3}{2 \left(r^2\,e^{-\frac{r_0}{r}}+2a\right)}|_{r=r_0}+\frac{r^3}{2 \left(r^2\,e^{-\frac{r_0}{r}}-2a\right)}|_{r=r_*^-}.
\end{equation}
As a special case when $a=0$, it follows that the two points coincide i.e., $r_*^+=r_*^-=r_0$, hence the shadow radius for the static wormhole given by Eq. (32). If we use the correspondence between the quasinormal mode and the conserved quantities along geodesics, for the corresponding prograde and retrograde modes, we can write
\begin{equation}
\omega_{\Re}^{\pm} = \frac{1}{R_s^{\pm}}\left(l+\frac{1}{2} \right).
\end{equation}

Alternatively, we can use Eq. (60) along with the last equation  to express the real part of the corresponding prograde/retrograde modes as
\begin{equation}
\omega_{\Re}^{\pm} = \frac{\left(l+\frac{1}{2} \right)}{\frac{r^3}{\pm r^2\,e^{-\frac{r_0}{r}}+2a}|_{r=r_*^{\pm}}}.
\end{equation}

\begin{table}[tbp]
\begin{tabular}{|l|l|l|l|l|l|}
\hline
\multicolumn{1}{|c|}{ } &  \multicolumn{1}{c|}{  $\mathrm{l}=1$ } & \multicolumn{1}{c|}{  $\mathrm{l}=1$ } & \multicolumn{1}{c|}{}\\\hline
  $a$ &\,\,\,\,$\omega_{\Re}^{+}$  &\,\,\,\,$\omega_{\Re}^{-}$  & \,\,\,\,$\bar{R}_s$   \\ \hline
0.0 & 0.551819 & -0.5518191   & 2.718281 \\ 
0.1 & 0.851819 & -0.4291367   & 2.628163 \\ 
0.2 & 1.151819 & -0.3799017 & 2.625338 \\
0.3 & 1.451819 & -0.3484413 &  2.669035  \\ 
0.4 & 1.751819 & -0.3254748 &  2.732452 \\ 
0.5 & 2.051819 & -0.3075117 &  2.804460  \\
0.6 & 2.351819 & -0.2928452 & 2.879981    \\
0.7 & 2.651819 & -0.2804967 &  2.956519  \\
0.8 & 2.951819 & -0.2699092 & 3.032792 \\\hline
\end{tabular}
\caption{The real part of QNMs and the typical shadow radius for a wormhole model $\Phi(r)=-r_0/r$ by varying the angular parameter $a$. Here we choose $r_0=1$.}
\end{table}

Finally, it may be useful to define yet another quantity, say the typical real part of QNMs
\begin{equation}
\bar{\omega}_{\Re}=\frac{1}{2}(\omega_{\Re}^+-\omega_{\Re}^-),
\end{equation}
which then can be used to find the typical shadow radius 
\begin{equation}\label{Eq52}
\bar{\omega}_{\Re} =  \frac{\left(l+\frac{1}{2} \right)}{\bar{R}_s}.
\end{equation}

In Fig. 5 (left panel) and Table V, we present the numerical values for the typical shadow radius and the corresponding prograde/retrograde mode. We see that, there is a reflecting point at some $a=a_c$, such that when $a<a_c$, the typical shadow radius decreases, and when $a>a_c$, the shadow radius increases. However there is reflecting point at some critical value $a_c$, such that if $a>a_c$ the shadow radius increases. In this particular model, for a very fast rotating wormholes we see that the shadow radius is bigger compared to the static case. 

 \subsection{ Model with $\Phi(r)=-\frac{r_0}{r}-\frac{r_0^2}{r^2}$}
\begin{figure*}
\includegraphics[width=8.4cm]{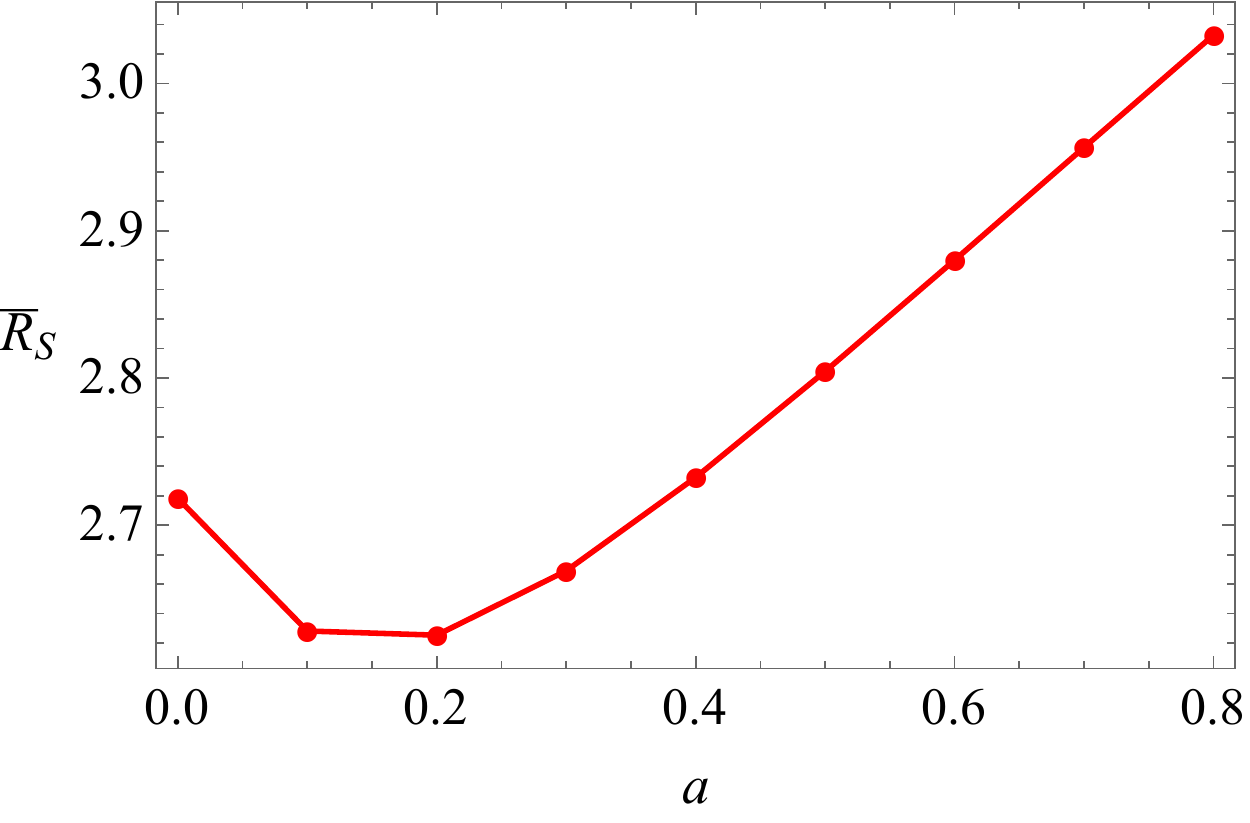}
\includegraphics[width=8.4cm]{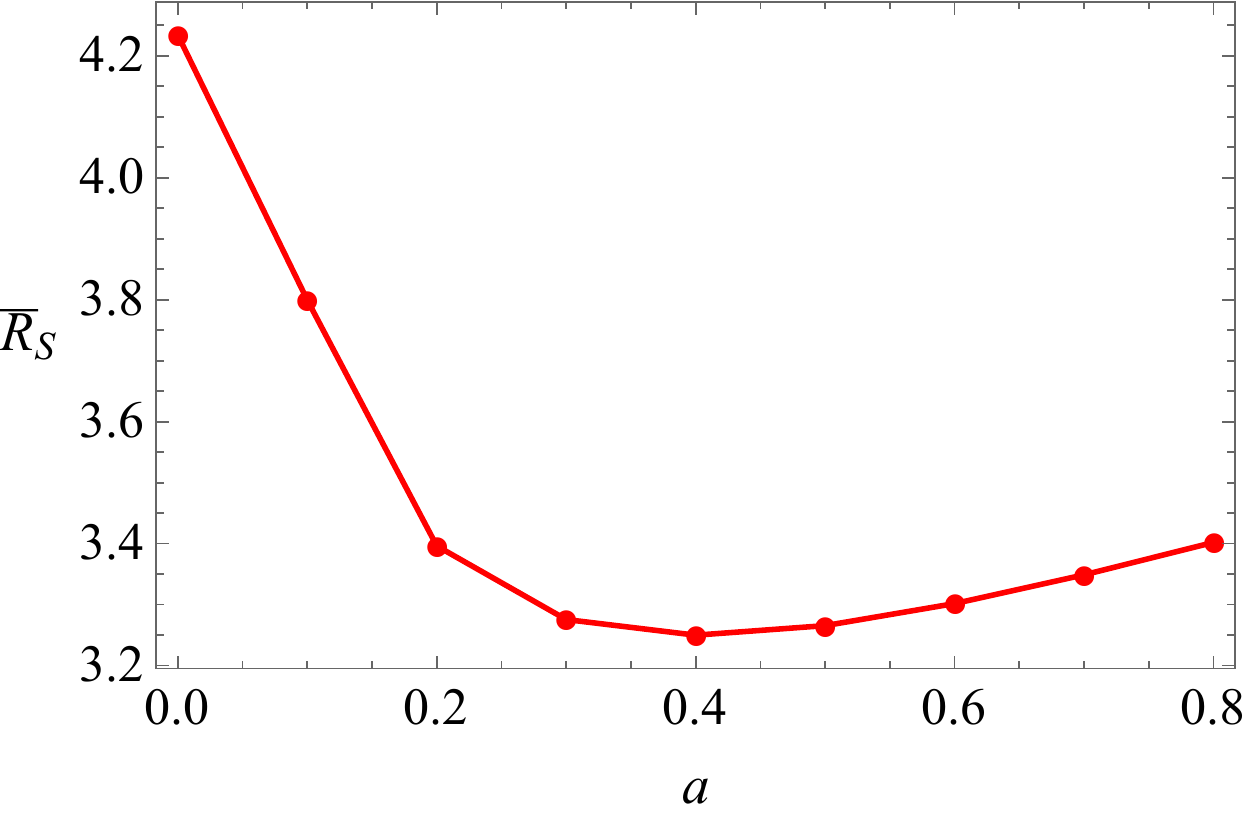}
\caption{Left panel: The typical shadow radius by varying $a$ for the model $\Phi(r)=-\frac{r_0}{r}$. Right panel: The typical shadow radius by varying $a$ for the model $\Phi(r)=-\frac{r_0}{r}-\frac{r_0^2}{r^2}$. Note that the viewing angle is $\theta_0=\pi/2$ and $r_0=1$ in both cases.}
\end{figure*}
In this particular wormhole model from Eq. (58) we obtain the following result
\begin{equation}\label{Eq34}
R_s^{\pm}=\frac{r^3}{\pm r^2\,e^{-\frac{r_0(r+r_0)}{r^2}}+2a}|_{r=r_*^{\pm}},
\end{equation}
where $r_*^{\pm}$ is determined from the second condition given by Eq. (59) yielding
\begin{eqnarray}\notag
&r|_{r=r_*^{\pm}}&-\frac{{R_s^{\pm}}^2 r_0}{r^2 e^{ \frac{2r_0}{r}} e^{ \frac{2r_0^2}{r^2}}}|_{r=r_*^{\pm}}-\frac{2 {R_s^{\pm}}^2 r_0^2}{r^3 e^{ \frac{2r_0}{r}} e^{ \frac{2r_0^2}{r^2}}}|_{r=r_*^{\pm}}\\
&+& \frac{2 a R_s^{\pm}}{r^2}|_{r=r_*^{\pm}}-\frac{8 a^2 {R_s^{\pm}}^2}{r^5}|_{r=r_*^{\pm}}=0.
\end{eqnarray}

We find that there is a contribution of the branch $R_s^-$ at the the outer light ring having $r_*^- > r_0$. Therefore, the typical shadow radius yields
\begin{equation}
\bar{R}_s=\frac{r^3}{ 2\left(r^2\,e^{-\frac{r_0(r+r_0)}{r^2}}+2a\right)}|_{r=r_0}+\frac{r^3}{2\left( r^2\,e^{-\frac{r_0(r+r_0)}{r^2}}-2a \right)}|_{r=r_+^-}
\end{equation}
As a special case when $a=0$, we have $r_*=r_{ph}=2r_0$, yielding Eq. (34). Finally we can obtain the real part of the corresponding prograde/retrograde modes as
\begin{equation}
\omega_{\Re}^{\pm} = \frac{\left(l+\frac{1}{2} \right)}{\frac{r^3}{\pm r^2\,e^{-\frac{r_0(r+r_0)}{r^2}}+2a}|_{r=r_*^{\pm}}}.
\end{equation}

\begin{table}[tbp]
\begin{tabular}{|l|l|l|l|l|l|}
\hline
\multicolumn{1}{|c|}{ } &  \multicolumn{1}{c|}{  $l=1$ } & \multicolumn{1}{c|}{  $l=1$ } & \multicolumn{1}{c|}{}\\\hline
  $a$ &\,\,\,\,$\omega_{\Re}^{+}$  &\,\,\,\,$\omega_{\Re}^{-}$  & \,\,\,\,$\bar{R}_s$   \\ \hline
0.0 & 0.354274 & -0.354274   & 4.234000 \\ 
0.1 & 0.503002 & -0.324898   & 3.799457 \\ 
0.2 & 0.803002 & -0.304607 & 3.396181  \\
0.3 & 1.103002 & -0.288960 &  3.275475  \\ 
0.4 & 1.403002 & -0.276216  &  3.249833 \\ 
0.5 & 1.703002 & -0.265477 & 3.265491  \\
0.6 & 2.003002 &  -0.256215 & 3.301661   \\
0.7 & 2.303002 & -0.248086 &  3.348805  \\
0.8 & 2.603002 & -0.240854 &  3.402035  \\\hline
\end{tabular}
\caption{The real part of QNMs and the typical shadow radius for a wormhole model $\Phi(r)=-r_0/r-r_0^2/r^2$ by varying the angular parameter $a$. Here we choose $r_0=1$.}
\end{table}
In Fig. 5 (right panel) and Table VI, we present the numerical values for the typical shadow radius and the corresponding prograde/retrograde mode. Similarly, we find a reflecting point at some $a=a_c$, such that when $a<a_c$, the typical shadow radius decreases, and when $a>a_c$, the shadow radius increases. In this particular model, for fast rotating wormholes we see that the shadow radius is smaller compared to the static case. 

%\begin{widetext}

\section{Conclusions}
In this paper we have studied the connection between the real part of quasinormal modes (QNMs) and the shadow radius in a wormhole spacetime. Firstly, we have investigated this correspondence in a static and spherically symmetric wormhole spacetime. To this end, we have chosen four different wormhole models having specific redshift functions. We have obtained the values for the real part of the QNMs in terms of the shadow radius. In general, we have shown that the values of QNMs decreases as we increase the wormhole throat radius. It is interesting that although this correspondence is accurate in the limit of large $l$, it works well also in the limit of small $l$.  
Secondly, we generalize this correspondence to the rotation wormhole spacetime and calculate the typical shadow radius of the rotating wormhole when viewed from the equatorial plane. We argue that due to the rotation, and depending on the specific model, the typical shadow radius can be smaller or bigger compared to the static case. In addition, we showed that there is a reflecting point at some value $a_c$. 
Thirdly, we have consider the possibility of having a traversable wormhole in the galactic center M87. Using the 
recent data reported by the EHT, we have constraint the wormhole throat radius $r_0$. For the model $\Phi(r)=-r_0/r$,  within $2\sigma$ confidence, we have found the following interval for the wormhole throat radius $r_0 \sim (4.6-8.1) \times 10^{-4}$ pc. 
In the second model having $\Phi(r)=-r_0/r-r_0^2/r^2$, we have found the interval $r_0 \sim (2.9-5.2) \times 10^{-4}$ pc, within $2\sigma$ confidence. Whether there is a wormhole in the galactic center is an open question, our results suggest that a wormhole can in principle mimic the black hole in the galactic center however more work is needed to distinguish these two objects.
%%%%%%%%%%%%%%%%%%%%%%%%%%%%%%%


\begin{thebibliography}{99}

\bibitem{flamm} L. Flamm,  Phys. Z. 17, 448 (1916)
\bibitem{Einstein35} A. Einstein and N. Rosen, Phys. Rev. \textbf{48}, 73-77 (1935).

\bibitem{Wheeler55} J. A. Wheeler, Phys. Rev. \textbf{97}, 511 (1955); R. W. Fuller and J. A. Wheeler, Phys. Rev. \textbf{128}, 919 (1962).

\bibitem{Thorne88} M. S. Morris and K. S. Thorne,  Am. J. Phys. \textbf{56}, 395 (1988); M. S. Morris K. S. Thorne and U. Yurtsever, Phys. Rev. D \textbf{61}, 1446 (1988).


\bibitem{teo} Edward Teo, Phys. Rev. D \textbf{58}, 024014 (1998).

\bibitem{Visser95} M. Visser, Lorentzian Wormholes: From Einstein to Hawking (American Institute of Physics, New York, 1995).



\bibitem{1} H. G. Ellis and J. Math. Phys. \textbf{14}, 104 (1973).

\bibitem{2} L. Chetouani and G. Clement, Gen. Rel. Grav. \textbf{16}, 111-119 (1984).  

\bibitem{3} N. Tsukamoto and T. Harada, Phys. Rev. D \textbf{95}, 024030 (2017).

\bibitem{4} K. Nakajima, H. Asada and Phys.Rev.D \textbf{85}, 107501 (2012).

\bibitem{5} A. Bhattachary, A. Potapov, Mod. Phys. Lett. A \textbf{25}, 2399 (2010). 

\bibitem{6} F. Abe, ApJ \textbf{725} (2010) 787-793

\bibitem{7} T. K. Dey and S. Sen, Mod. Phys. Lett. A \textbf{23}, 953-962, (2008).

\bibitem{8}R.~Shaikh and S.~Kar,
  %``Gravitational lensing by scalar-tensor wormholes and the energy conditions,''
  Phys.\ Rev.\ D {\bf 96}, no. 4, 044037 (2017).
  
\bibitem{9} K. Jusufi, Int. J. Geom. Methods Mod. Phys.  14 (2017) 1750179.


%\cite{Dai:2019mse}
\bibitem{10}
D.~C.~Dai and D.~Stojkovic,
%``Observing a Wormhole,''
Phys. Rev. D \textbf{100} (2019) no.8, 083513
%18 citations counted in INSPIRE as of 29 Jul 2020

%\cite{Simonetti:2020vhw}
\bibitem{11}
J.~H.~Simonetti, M.~J.~Kavic, D.~Minic, D.~Stojkovic and D.~C.~Dai,
%``A sensitive search for wormholes,''
[arXiv:2007.12184 [gr-qc]].
%0 citations counted in INSPIRE as of 29 Jul 2020


  %\cite{Synge66}
  \bibitem{Synge66} 
  J.~L.~Synge,
  %``The Escape of Photons from Gravitationally Intense Stars,''
  Mon.\ Not.\ Roy.\ Astron.\ Soc.\  {\bf 131}, no. 3, 463 (1966).
  %doi:10.1093/mnras/131.3.463
  %%CITATION = doi:10.1093/mnras/131.3.463;%%
  %83 citations counted in INSPIRE as of 06 Nov 2019
 
   
  %\cite{Luminet79}
  \bibitem{Luminet79} 
  J.-P.~Luminet,
  %``Image of a spherical black hole with thin accretion disk,''
  Astron.\ Astrophys.\  {\bf 75}, 228 (1979).
  %%CITATION = AAEJA,75,228;%%
  %240 citations counted in INSPIRE as of 06 Nov 2019

  %\cite{DeWitt73}
  \bibitem{DeWitt73} 
  J. M. Bardeen, in Black Holes (Proceedings, Ecole d'Eté de Physique Théorique: Les Astres Occlus : Les Houches, France, August, 1972) edited by C.~DeWitt and B.~S.~DeWitt
  %``Proceedings, Ecole d'Eté de Physique Théorique: Les Astres Occlus : Les Houches, France, August, 1972,''
  %%CITATION = INSPIRE-1357064;%%
  
  
\bibitem{w1} C. Bambi, Phys. Rev. D 87, 107501 (2013) 
  %\cite{Shaikh:2018kfv}
  
 \bibitem{w2} T. Ohgami and N. Sakai, Wormhole shadows, Phys. Rev. D 91, 124020 (2015)
 
\bibitem{w3} P. G. Nedkova, V. Tinchev, and S. S. Yazadjiev, Phys. Rev. D 88, 124019 (2013).
\bibitem{w4}
R.~Shaikh,
%``Shadows of rotating wormholes,''
Phys. Rev. D \textbf{98} (2018) no.2, 024044
[arXiv:1803.11422 [gr-qc]].
%63 citations counted in INSPIRE as of 29 Jul 2020

%\cite{Gyulchev:2018fmd}
\bibitem{w5}
G.~Gyulchev, P.~Nedkova, V.~Tinchev and S.~Yazadjiev,
%``On the shadow of rotating traversable wormholes,''
Eur. Phys. J. C \textbf{78} (2018) no.7, 544
[arXiv:1805.11591 [gr-qc]].
%30 citations counted in INSPIRE as of 29 Jul 2020
  
  %\cite{Amir:2018pcu}
\bibitem{w6}
M.~Amir, K.~Jusufi, A.~Banerjee and S.~Hansraj,
%``Shadow images of Kerr-like wormholes,''
Class. Quant. Grav. \textbf{36} (2019) no.21, 215007
[arXiv:1806.07782 [gr-qc]].
%29 citations counted in INSPIRE as of 29 Jul 2020

\bibitem{Akiyama1} K. Akiyama et al. (Event Horizon Telescope), Astrophys.
J. 875 (2019) L1.

\bibitem{Akiyama2} K. Akiyama et al. (Event Horizon Telescope), Astrophys.
J. 875 (2019) L4.



\bibitem{BertiCardosoWill} E. Berti, V. Cardoso and C. Will, Phys. Rev. D 73
(2006) 064030.

\bibitem{Regge} T. Regge and J. A. Wheeler, Phys. Rev. 108 (1957) 1063.

\bibitem{Zerilli} F. J. Zerilli, Phys. Rev. D 2 (1970) 2141.

  
  %\cite{Berti:2005eb}
\bibitem{111}
  E.~Berti and K.~D.~Kokkotas,
  %``Quasinormal modes of Kerr-Newman black holes: Coupling of electromagnetic and gravitational perturbations,''
  Phys.\ Rev.\ D {\bf 71} (2005) 124008
  [gr-qc/0502065].
  %%CITATION = doi:10.1103/PhysRevD.71.124008;%%
  %80 citations counted in INSPIRE as of 06 Apr 2020
  
  \bibitem{222} B. Mashhoon, Phys. Rev. D 31, 290 (1985)
  
  %\cite{Konoplya:2011qq}
\bibitem{333}
  R.~A.~Konoplya and A.~Zhidenko,
  %``Quasinormal modes of black holes: From astrophysics to string theory,''
  Rev.\ Mod.\ Phys.\  {\bf 83} (2011) 793
  %%CITATION = doi:10.1103/RevModPhys.83.793;%%
  %504 citations counted in INSPIRE as of 08 Apr 2020

\bibitem{444} V. Ferrari and B. Mashhoon, Phys. Rev. D 30 (1984) 295.

\bibitem{555} B. F. Schutz and C. M. Will, Astrophys. J. Lett. 291 (1985)
L33.

\bibitem{666} S. Iyer and C. M. Will, Phys. Rev. D 35 (1987) 3621.

\bibitem{777} R. A. Konoplya, Phys. Rev. D 68 (2003) 024018.


\bibitem{Konoplya:2018ala}
  R.~A.~Konoplya,
  %``How to tell the shape of a wormhole by its quasinormal modes,''
  Phys.\ Lett.\ B {\bf 784} (2018) 43
  [arXiv:1805.04718 [gr-qc]].
  %%CITATION = doi:10.1016/j.physletb.2018.07.025;%%
  %17 citations counted in INSPIRE as of 24 Jan 2020
 

  %\cite{Churilova:2019qph}
\bibitem{Churilova:2019qph}
M.~S.~Churilova, R.~A.~Konoplya and A.~Zhidenko,
%``Arbitrarily long-lived quasinormal modes in a wormhole background,''
Phys. Lett. B \textbf{802} (2020), 135207
[arXiv:1911.05246 [gr-qc]].
%6 citations counted in INSPIRE as of 29 Jul 2020

%\cite{Oliveira:2018oha}
\bibitem{Oliveira:2018oha}
R.~Oliveira, D.~M.~Dantas, V.~Santos and C.~A.~S.~Almeida,
%``Quasinormal modes of bumblebee wormhole,''
Class. Quant. Grav. \textbf{36} (2019) no.10, 105013
[arXiv:1812.01798 [gr-qc]].
%9 citations counted in INSPIRE as of 29 Jul 2020
%\cite{Dent:2020nfa}

 \bibitem{AbbottBH} B. P. Abbott et al. (LIGO Scientific and Virgo
Collaborations), Phys. Rev. Lett. 116 (2016) 061102.


\bibitem{gw}
J.~B.~Dent, W.~E.~Gabella, K.~Holley-Bockelmann and T.~W.~Kephart,
%``The Sound of Clearing the Throat: Gravitational Waves from a Black Hole Orbiting in a Wormhole Geometry,''
[arXiv:2007.09135 [gr-qc]].
%1 citations counted in INSPIRE as of 29 Jul 2020

 \bibitem{cardoso} V. Cardoso, A. S. Miranda, E. Berti, H. Witek, and V. T. Zanchin, Phys. Rev. D 79, 064016 (2009)
  %\cite{Stefanov:2010xz}

%\cite{Hod:2017xkz}
\bibitem{Hod:2017xkz}
  S.~Hod,
  %``Upper bound on the radii of black-hole photonspheres,''
  Phys.\ Lett.\ B {\bf 727} (2013) 345
  %%CITATION = doi:10.1016/j.physletb.2013.10.047;%%
  %28 citations counted in INSPIRE as of 08 Apr 2020
  
\bibitem{Wei:2019jve}
  S.~W.~Wei and Y.~X.~Liu,
  %``Null Geodesics, Quasinormal Modes, and Thermodynamic Phase Transition for Charged Black Holes in Asymptotically Flat and dS Spacetimes,''
  arXiv:1909.11911 [gr-qc].
  %%CITATION = ARXIV:1909.11911;%%
  %1 citations counted in INSPIRE as of 22 Dec 2019
  
  
  \bibitem{Stefanov:2010xz}
  I.~Z.~Stefanov, S.~S.~Yazadjiev and G.~G.~Gyulchev,
  %``Connection between Black-Hole Quasinormal Modes and Lensing in the Strong Deflection Limit,''
  Phys.\ Rev.\ Lett.\  {\bf 104} (2010) 251103
  %%CITATION = doi:10.1103/PhysRevLett.104.251103;%%
  %46 citations counted in INSPIRE as of 19 Dec 2019
  
 \bibitem{Konoplya:2017wot}
  R.~A.~Konoplya and Z.~Stuchlík,
  %``Are eikonal quasinormal modes linked to the unstable circular null geodesics?,''
  Phys.\ Lett.\ B {\bf 771} (2017) 597
  %%CITATION = doi:10.1016/j.physletb.2017.06.015;%%
  %57 citations counted in INSPIRE as of 03 Feb 2020
  
  %\cite{Jusufi:2019ltj}
\bibitem{Jusufi:2019ltj}
  K.~Jusufi,
  %``Quasinormal Modes of Black holes Surrounded by Dark Matter and Their Connection with Shadow Radius,''
 Phys. Rev. D  {\bf 101}, 084055 (2020)
  %%CITATION = ARXIV:1912.13320;%%
  %1 citations counted in INSPIRE as of 22 Feb 2020
  %\cite{Feng:2019zzn}
  %\cite{Liu:2020ola}


%\cite{Jusufi:2020dhz}
\bibitem{Jusufi:2020dhz}
K.~Jusufi,
%``Connection Between the Shadow Radius and Quasinormal Modes in Rotating Spacetimes,''
Phys. Rev. D {\bf 101}, 124063 (2020)
%4 citations counted in INSPIRE as of 27 Jun 2020

\bibitem{Liu:2020ola}
  C.~Liu, T.~Zhu, Q.~Wu, K.~Jusufi, M.~Jamil, M.~Azreg-Aïnou and A.~Wang,
  %``Shadow and Quasinormal Modes of a Rotating Loop Quantum Black Hole,''
  Phys.\ Rev.\ D {\bf 101} (2020) no.8,  084001
  %%CITATION = doi:10.1103/PhysRevD.101.084001;%%
  %3 citations counted in INSPIRE as of 07 Apr 2020 
  
  %\cite{Amir:2018pcu}
\bibitem{Amir:2018pcu}
M.~Amir, K.~Jusufi, A.~Banerjee and S.~Hansraj,
%``Shadow images of Kerr-like wormholes,''
Class. Quant. Grav. \textbf{36} (2019) no.21, 215007
[arXiv:1806.07782 [gr-qc]].
%29 citations counted in INSPIRE as of 29 Jul 2020

%\cite{Hendi:2020knv}
\bibitem{hendi}
  S.~H.~Hendi, S.~N.~Sajadi and M.~Khademi,
  %``Physical Properties of a Regular Rotating Black Hole: Thermodynamics, Stability, Quasinormal Modes,''
  arXiv:2006.11575 [gr-qc].
  %%CITATION = ARXIV:2006.11575;%%
  
%\cite{Cuadros-Melgar:2020kqn}
\bibitem{Cuadros-Melgar:2020kqn}
B.~Cuadros-Melgar, R.~D.~B.~Fontana and J.~de Oliveira,
%``Analytical correspondence between shadow radius and black hole quasinormal frequencies,''
[arXiv:2005.09761 [gr-qc]].
%3 citations counted in INSPIRE as of 29 Jul 2020

%\cite{Guo:2020nci}
\bibitem{Guo:2020nci}
Y.~Guo and Y.~G.~Miao,
%``Null geodesics, quasinormal modes and the correspondence with shadows in high-dimensional Einstein-Yang-Mills spacetimes,''
[arXiv:2007.08227 [hep-th]].
%0 citations counted in INSPIRE as of 29 Jul 2020
  
\bibitem{Feng:2019zzn}
  X.~H.~Feng and H.~Lu,
  %``On the Size of Rotating Black Holes,''
  arXiv:1911.12368 [gr-qc].
  %%CITATION = ARXIV:1911.12368;%%
  %4 citations counted in INSPIRE as of 22 Feb 2020
  %\cite{Yang:2012he}
 


%\cite{Allahyari:2019jqz}
\bibitem{Allahyari:2019jqz}
A.~Allahyari, M.~Khodadi, S.~Vagnozzi and D.~F.~Mota,
%``Magnetically charged black holes from non-linear electrodynamics and the Event Horizon Telescope,''
JCAP \textbf{02} (2020), 003
[arXiv:1912.08231 [gr-qc]].
%31 citations counted in INSPIRE as of 29 Jul 2020

%\cite{Khodadi:2020jij}
\bibitem{Khodadi:2020jij}
M.~Khodadi, A.~Allahyari, S.~Vagnozzi and D.~F.~Mota,
%``Black holes with scalar hair in light of the Event Horizon Telescope,''
[arXiv:2005.05992 [gr-qc]].
%7 citations counted in INSPIRE as of 30 Jul 2020

%\cite{Bronnikov:2018nub}
\bibitem{Bronnikov:2018nub}
K.~A.~Bronnikov and K.~A.~Baleevskikh,
%``On gravitational lensing by symmetric and asymmetric wormholes,''
Grav. Cosmol. \textbf{25}, no.1, 44-49 (2019)

\end{thebibliography}
\end{document}